# Efficient resonance management in ultrahigh-Q one-dimensional photonic crystal nanocavities fabricated on 300 mm SOI CMOS platform

*Weiqiang Xie,* [1,*] *Peter Verheyen* [2], *Marianna Pantouvaki* [2], *Joris Van Campenhout,* [2] *and Dries Van Thourhout* [1,*]

E-mail: jdxwq@hotmail.com, Dries.VanThourhout@UGent.be

[1] Photonics Research Group, INTEC Department, Ghent University-imec, Technologiepark-Zwijnaarde 126, 9052 Gent, Belgium
[2] imec, Kapeldreef 75, 3001 Leuven, Belgium

Photonic crystal (PhC) nanocavities have demonstrated unique capabilities in terms of light confinement and manipulation. As such, they are becoming attractive for the design of novel resonance-based photonic integrated circuits (PICs). Here two essential challenges arise however − how to realize ultrahigh-$Q$ PhC cavities using standard fabrication processes compatible with large volume fabrication, and how to efficiently integrate them with other standard building blocks, available in exiting PIC platforms. In this work, we demonstrate ultrahigh-$Q$ 1D PhC nanocavities fabricated on a 300 mm SOI wafer by optical lithography, with a record $Q$ factor of up to 0.84 million. Moreover, we show efficient mode management in those oxide embedded cavities by coupling them with an access waveguide and realize two critical components: notch filters and narrow-band reflectors. In particular, they allow both single-wavelength and multi-wavelength operation, at the desired resonant wavelengths, while suppressing all other wavelengths over a broad wavelength range (>100 nm). Compared to traditional cavities, this offers a fantastic strategy for implementing resonances precisely in PIC designs with more freedom in terms of wavelength selectivity and the control of mode number. Given their compatibility with optical lithography and compact footprint, the realized 1D PhC nanocavities will be of profound significance for designing compact and novel resonance-based photonic components on large scale.

## 1. Introduction

In silicon photonics, an optical cavity is one of the key building blocks for both passive and active photonic integrated circuits (PICs). Photonic crystal (PhC) cavities, using a photonic bandgap to confine light, have gained lasting interest in Si photonics over the past two decades [1-7], since they enable both ultrahigh quality factors ($Q$) (up to millions) and ultrasmall cavity modal volumes ($V$) (at the scale of a cubic wavelength), benefiting from the high optical index and low loss of crystalline Si. Thus, PhC nanocavities offer more advanced photon manipulation in terms of lifetime and density at a desired resonant wavelength, compared to traditional cavities such as ring resonators, Fabry–Pérot cavities, and distributed feedback resonators. Therefore, they have found applications in many areas including ultralow threshold lasers [8, 9], compact optical switches and modulators [10-12], ultrasmall filters [13-15], and slow light [16, 17]. With a continuous improvement in design and fabrication we have witnessed an increase of the $Q$ factor from a few hundred for the earliest PhC nanocavities [1] to values on the order of millions for most recent devices, with a record $Q$ of 11 million [7] and 0.72 million [6], in two-dimensional (2D) and one-dimensional (1D) air-bridge Si PhC nanocavities respectively. However, given their subwavelength feature size, most demonstrated PhC nanocavities relied on high-resolution electron beam lithography (EBL), limiting their scalability to large volume applications. Therefore, recently several groups worked on the development of PhC cavities defined using optical lithography [18-20]. $Q$ factors up to 2 million and 0.6 million have been achieved for air-bridge and oxide-embedded 2D PhC nanocavities respectively. For oxide-embedded 1D PhC cavities, the highest $Q$ factor demonstrated thus far is limited to 0.04 million [20]. Compared with 2D devices, 1D PhC nanocavities have a relatively larger sidewall surface, which imposes more rigorous requirements on limiting fabrication imperfections such as lithography distortion and etching roughness. Hence, the realization of ultrahigh $Q$ factors in 1D PhC cavities defined by optical lithography remains a challenge.

Furthermore, with the complexity of silicon PICs steadily increasing, there is a great interest in integrating those PhC cavities in larger scale optical circuits. This calls for an efficient and flexible approach for combining PhC cavities with existing building blocks and in particular with silicon wire waveguides. Taking these considerations into account, the 1D PhC cavity has several advantages over its 2D counterpart. First of all, it utilizes a 1D photonic bandgap to confine light along the direction of propagation in a periodic waveguide structure, light being confined in the other directions by total internal reflection (index-guiding), while the 2D cavity needs a 2D PhC structure to confine light in plane, which significantly increases the device footprint. Secondly, because they are built on waveguide wires, the 1D PhC cavities can be directly coupled with an inline waveguide or side coupled to a bus waveguide, potentially allowing for simple and effective integration and facilitating the connection with other components in a compact layout. In contrast, a specially designed 2D PhC waveguide is necessary for coupling with a 2D PhC cavity, which further increases the device footprint. Moreover, the transition between a 2D PhC waveguide and a regular waveguide wire requires careful design [21] to avoid introducing considerable loss. This is a further motivation for developing ultrahigh-$Q$ 1D PhC cavities as well as the related building blocks on a CMOS compatible platform.

In this work, we report the realization of ultrahigh-$Q$ 1D PhC nanocavities fabricated on a 300 mm SOI wafer in a CMOS pilot line using 193 nm immersion lithography. $Q$ factors of up to 0.84 million are achieved for oxide-embedded 1D PhC cavities, which are, to the best of our knowledge, the highest values reported for 1D Si PhC nanocavities (both air and oxide cladded). Moreover, in both simulation and experiment, we show efficient mode management in simple waveguide-coupled 1D PhC nanocavities. We demonstrate novel resonance-based narrow-band notch filters and reflectors with bandwidth comparable to the linewidth of the resonance mode of the cavity (~1 GHz at 1550 nm), while their footprint is only 4μm × 22μm. Both reflectors and transmission filters enable single-wavelength operation over a broad wavelength band (>100 nm), and they also allow multi-wavelength operation at the desired resonances by coupling a single bus waveguide with multiple PhC nanocavities with arbitrary yet predefined resonance frequencies. Given their fabrication using optical lithography techniques and the fact they are oxide embedded, together with their straightforward coupling with waveguide wires and compact footprint, the demonstrated cavities can readily be integrated in more complex passive or even active photonic ICs.

## 2. Numerical Design and Theoretical Analysis

Fig. 1 schematically shows the structure of our 1D PhC nanocavity. The starting point is an SOI substrate with a 2 µm $SiO_2$ box layer and 220 nm Si layer. The cavity consists of a periodic array of circular holes along the waveguide and the entire structure is embedded in the oxide with planarized top surface by our fabrication as shown in Fig. 1(b). The period of the PhC lattice, defined as the center-to-center distance between two holes, is denoted as *a*. The design of the cavity is based on the mode-gap modulation approach [6, 20]. Here, we modulate the radius of the PhC hole from the center symmetrically to the two sides of the cavity such that the mode gap of the 1D PhC waveguide is varied. As a result, in the modulated region a cavity mode (defect mode) can be created in the lower-lying dielectric band of the PhC waveguide, which is then confined within the bandgap of the two side PhC mirrors. In a 1D nanocavity with enough mirror holes (thus no guided mode loss at both sides), the cavity *Q* is limited by intrinsic radiative loss due to the incomplete bandgap of the 1D PhC waveguide. A gradual modulation scheme as applied here can reduce the radiative loss dramatically, resulting in an ultrahigh *Q*, at least in simulations. In practice, the *Q* factor is also limited by roughness and imperfection induced scattering loss, which actually is a dominant contribution to the total intrinsic loss. We define the intrinsic *Q* factor in a 1D PhC cavity with infinitely extended mirrors as

$$1/Q_{int} = 1/Q_{rad} + 1/Q_{scat}, \tag{1}$$

where $Q_{int}$ is the intrinsic quality factor, $Q_{rad}$ is the quality factor due to radiative loss into free space, and $Q_{scat}$ is the scattering loss related quality factor. Note that the material loss is neglected here.

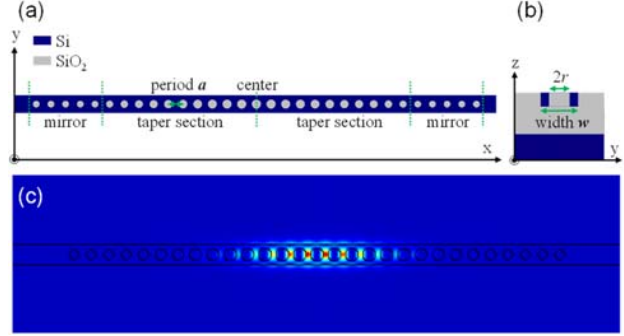

**Figure 1.** Design of 1D PhC nanocavity. (a) Schematics of 1D SOI PhC nanocavity with circular holes embedded in silicon oxide matrix, consisting of the taper and mirror sections. The whole structure is symmetric with respect to the cavity center. (b) Cross-sectional view of the cavity. (c) Simulated mode profile ($|E|^2$) of the fundamental cavity mode.

We analyzed the cavity mode numerically with the three-dimensional finite-difference time-domain (FDTD) method, using a freely available software package [22], to find out the optimal cavity parameters aiming for ultrahigh-$Q$ resonance modes in the C-band. FDTD simulation can also directly calculate $Q_{rad}$ for the cavity with adequate mirrors (e.g. ≥ 20). The final design of the cavity has a lattice constant of $a$=365 nm, a radius of $r_0$=0.3$a$ for the center hole and 0.22$a$ for the mirror holes, and a waveguide width of $w_0$=1.2$a$. From the simulation, the photonic bandgap of the mirror ranges from a wavelength of 1370 nm up to 1600 nm. The radius of the holes between the center and the mirror is parabolically tapered down from 0.30$a$ to 0.22$a$ with a total of 10 holes in the tapered section, as shown in Fig. 1(a). From FDTD simulation, we calculated quality factors of up to $10^7$ at the resonance wavelength of around 1550 nm for a design with 20 mirror holes at each side. The footprint of the cavity with such $Q$ is only 0.5μm × 22μm. Fig. 1(b) shows the simulated field profile for the fundamental mode of the cavity, a transverse-electric (TE) like mode. The mode is tightly localized around the center of the cavity with a modal volume on the order of $(\lambda/n)^3$. In the actual mask design used for fabrication, both the radius of the cavity hole and the PhC waveguide width are slightly scaled, to tune the resonance wavelength when needed and to compensate any possible bias from fabrication.

To access the cavity mode, we first utilize an inline waveguide-cavity configuration where the two sides of the cavity are connected to an inline waveguide. In this case, the mirror section can have a different number of holes to tune the radiation strength of the cavity mode through the mirror, and this radiated light is coupled to the waveguide mode. Thus, the total $Q$ factor (i.e., loaded $Q$ factor) for an inline PhC cavity with finite length mirrors can be expressed as

$$1/Q_{\text{tot}} = 1/Q_{\text{int}} + 1/Q_{\text{w}}, \qquad (2)$$

where $Q_{\text{w}}$ is the waveguide-cavity coupling related quality factor. With altering the number of holes in the PhC mirror, $Q_{\text{w}}$ can be tuned correspondingly. For the inline PhC cavity, the maximum transmission, at the resonance wavelength $\lambda_{\text{res}}$, is given by $T(\lambda_{\text{res}}) = (Q_{\text{tot}}/Q_{\text{w}})^2$ [23]. Combining this equation with Eq. (2), we can extract $Q_{\text{int}}$ as well as $Q_{\text{w}}$ directly from the transmission spectrum (which gives us directly $Q_{\text{tot}}$ and $T$). Note that the FDTD simulation for cavities with finite length mirrors give us the total radiative loss, i.e., the sum of $Q_{\text{rad}}$ and $Q_{\text{w}}$.

Inline PhC cavities are very useful for realizing narrow-band filters. If the waveguide-coupling loss is dominant in a cavity (e.g., $Q_{\text{int}} \gg Q_{\text{w}}$), the bandwidth is determined by the designed $Q_{\text{w}}$, and the transmission at resonance is ~100%. Therefore, a high $Q_{\text{int}}$ is pursued in practice. Besides inline cavities, we also considered the bus waveguide-coupled cavity structure, as shown in Fig. 2(a), which has the potential to be used both in transmission and reflection. Instead of using a straight bus waveguide parallel to the entire 1D PhC cavity, we implemented a curved bus waveguide, which incorporates a straight section in the center for coupling with the cavity. As the fundamental mode of the cavity is confined in the center of the cavity as indicated in Fig. 1(c), our waveguide design not only ensures effective control of coupling strength but also minimizes any parasitic losses due to the bus waveguide. Besides, this design could potentially suppress the coupling of high-order cavity modes, which exhibit less spatial overlap with the waveguide mode. In the fabricated devices, the bus waveguide has a straight section of 2.5 µm and then is bent out at both ports with a bending radius of 3.0 µm. The width of the bus waveguide is fixed at 450 nm and the gap between the waveguide and the cavity is varied from 250 nm to 450 nm to tune the coupling strength.

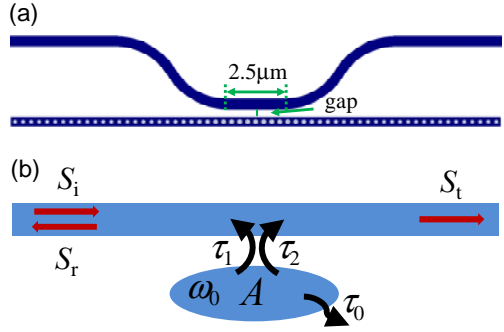

**Figure 2.** Side-coupling configuration of 1D PhC cavity and bus waveguide. (a) Design of 1D PhC nanocavity laterally coupled with a bus waveguide. The cavity has 20 mirror holes at each side. The structure is symmetric with respect to the center. (b) Schematics of side-coupling model between a single-mode waveguide and a standing-wave cavity with a single resonant mode.

We analyze the waveguide-coupled nanocavity with a side-coupling model [24] between a single-mode waveguide and a standing-wave cavity with a single resonant mode, as shown in Fig. 2(b). Here $A$ is the field amplitude of the cavity mode with a resonant frequency of $\omega_0$, with $S_i/S_r$ the incident/reflected field amplitudes at the input (left) waveguide port, and $S_t$ the transmitted field amplitude at the output (right) waveguide port. $1/\tau_0$ is the decay rate due to intrinsic cavity loss without the bus waveguide (thus intrinsic quality factor is $Q_{int}=\omega_0\tau_0/2$) while $1/\tau_1$ and $1/\tau_2$ are the decay rates to the waveguide in forwards and backwards direction, and associated with the quality factors $Q_1$ and $Q_2$ respectively. In our structure, by symmetry, we have $\tau_1=\tau_2=\tau_e$ ($Q_1=Q_2=Q_e=\omega_0\tau_e/2$). The amplitudes $S_i$, $S_r$, and $S_t$ are normalized so that their squared magnitude is the power in the waveguide mode. Similarly, the squared magnitude of $A$ is equal to the energy of the cavity mode.

When we launch a single waveguide mode $S_i$ at the input port, the incident light can be coupled to the cavity mode, and the light in the cavity is also coupled to the waveguide mode or decays as an intrinsic loss. According to coupled-mode theory [23, 24], the equations for the evolution of the cavity/waveguide mode amplitude in time is given by

$$\frac{dA}{dt} = i\omega_0 A - \left(\frac{1}{\tau_0} + \frac{1}{\tau_e} + \frac{1}{\tau_e}\right) A + \sqrt{\frac{2}{\tau_e}} S_i, \qquad (3)$$

$$S_r = \sqrt{\frac{2}{\tau_e}} A, \qquad (4)$$

$$S_t = S_i - \sqrt{\frac{2}{\tau_e}} A. \qquad (5)$$

The reflection and transmission are defined as

$$R = \frac{|S_r|^2}{|S_i|^2}, \quad T = \frac{|S_t|^2}{|S_i|^2}. \qquad (6)$$

For an input amplitude $S_i = e^{i\omega t}$, in the steady state, $dA/dt = i\omega A$ and we have

$$\sqrt{\frac{2}{\tau_e}} S_i = i(\omega - \omega_0) A + \left(\frac{1}{\tau_0} + \frac{2}{\tau_e}\right) A. \qquad (7)$$

Then, we obtain

$$R = \left|\frac{2/\tau_e}{1/\tau_0 + 2/\tau_e + i(\omega - \omega_0)}\right|^2, \quad T = \left|\frac{1/\tau_0 + i(\omega - \omega_0)}{1/\tau_0 + 2/\tau_e + i(\omega - \omega_0)}\right|^2. \qquad (8)$$

We define $\delta = (\omega - \omega_0)/\omega_0$ and $1/Q_w = 1/Q_e + 1/Q_e$. The total quality factor $Q_{tot}$ of the coupled cavity is given by

$$1/Q_{tot} = 1/Q_{int} + 1/Q_w. \qquad (9)$$

Then we obtain

$$R(\omega) = \frac{(1 - Q_{tot}/Q_{int})^2}{1 + 4Q_{tot}^2 \delta^2}, \quad T(\omega) = 1 - \frac{1 - (Q_{tot}/Q_{int})^2}{1 + 4Q_{tot}^2 \delta^2}. \qquad (10)$$

At resonance $\omega = \omega_0$, thus we have

$$R = (1 - Q_{tot}/Q_{int})^2, \quad T = (Q_{tot}/Q_{int})^2. \qquad (11)$$

From Eqs. (10) and (11), for a side waveguide-coupled nanocavity, at the cavity resonance the spectral response exhibits a Lorentzian shape with a bandwidth of $\omega_0/Q_{tot}$, and the peak intensity/depth at resonance is determined by the ratio of $Q_{tot}/Q_{int}$. Assuming an intrinsic $Q$ factor of $Q_{int} = 10^6$, we plot the reflection and transmission spectra in Fig. 3(a) and 3(b), respectively, with the peak values at resonance vs. $Q_{tot}$ shown in Fig. 3(c). It can be seen that the reflectance and transmission extinction ratio are strongly dependent on $Q_{tot}$, which can be simply controlled by changing the coupling gap. In the regime of $Q_{tot} \ll Q_{int}$ (i.e., $Q_w \ll Q_{int}$), e.g., $Q_{tot}=0.05Q_{int}$, the reflectance is more than 90% and the transmission is suppressed by up to 26 dB, while, if a narrow bandwidth is preferred, e.g., $Q_{tot}=0.3Q_{int}$, one can still attain a considerable reflectance of 50% and a transmission notch of 10 dB. In any case, a high intrinsic cavity quality factor $Q_{int}$ is essential to enable narrow-band high-efficiency reflection and transmission. A higher $Q_{int}$ also provides more freedom for tuning $Q_{tot}$, reflectance, and transmittance in practical device design. From the mode-coupling analysis, it is clear that a side-coupling scheme using a single bus waveguide provides effective resonance management for a 1D PhC nanocavity in which the resonance mode can be accessed by reflection and transmission coupled to the waveguide separately. By coupling a bus waveguide with multiple 1D nanocavities having different resonances, multi-wavelength operation can be achieved.

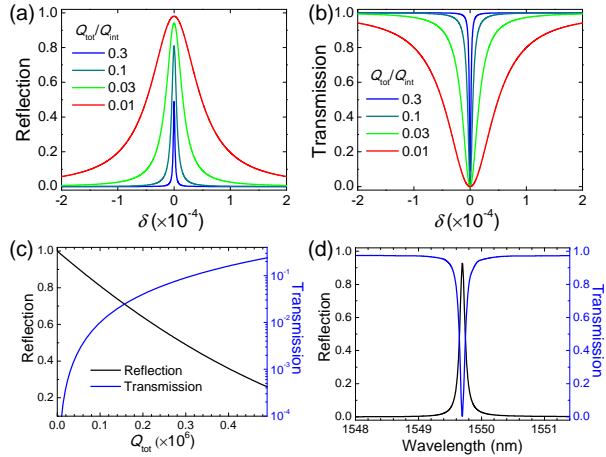

**Figure 3.** Analytical and numerical results of spectral response for a side waveguide-coupled 1D PhC cavity. (a) and (b) Analytical results of reflection and transmission spectra for different $Q_{tot}/Q_{int}$, with a cavity intrinsic Q factor of $Q_{int}=10^6$. (c) The reflectance and transmittance at resonance as a function of $Q_{tot}$. (d) FDTD simulated reflection and transmission spectra at the cavity fundamental resonance for the design as shown in Fig. 2(a). The coupling gap is 350 nm in simulation.

To verify our analysis, we also calculated the spectral response for the structure shown in Fig. 2 using a FDTD simulation and show the results in Fig. 3(d) for a coupling gap of 350 nm.

It is obvious that for a side waveguide-coupled 1D nanocavity the simulated spectral response is consistent with the analytical results from the coupled-mode theory as shown in Fig. 3(a) and 3(b). Besides, the FDTD simulation can give us guidance on the design of the coupling strength ($Q_w$) between bus waveguide and cavity. For instance, for a coupling gap of 350 nm, at the fundamental resonance, $Q_{tot}$ is ~$1.6\times10^4$, implying $Q_w \approx Q_{tot}$ (given $Q_{int} > 10^7$). By applying Eqs. (9) and (11), this, in turn, can give us the estimation of the transmission, reflection, and $Q_{tot}$ in a real fabricated device at a certain coupling gap once the cavity $Q_{int}$ is measured.

## 3. Fabrication and Measurement Results

We fabricated the 1D PhC nanocavities on a 300 mm SOI wafer in a CMOS pilot line using 193 nm immersion lithography [20]. The devices were oxide-embedded through oxide deposition and planarization down to the top of the silicon layer, and thus allow straightforward integration with other optical functions in the future. To allow direct optical characterization, the 1D PhC nanocavities were integrated with out-of-plane grating couplers optimized for TE polarization. Fig. 4 shows the scanning electron microscope (SEM) images of the fabricated inline and waveguide-coupled 1D nanocavities, indicating well-defined circular holes and the coupling region. The fabricated devices were characterized by measuring the transmission and reflection spectra, using a tunable laser (from 1500 to 1620 nm), and cleaved single-mode fibers for input/output coupling. The transmittance and reflectance were obtained by normalizing the spectra to a reference waveguide. In particular, for the reflection measurement, we employed index-matching fluid to suppress the reflection from fiber facets.

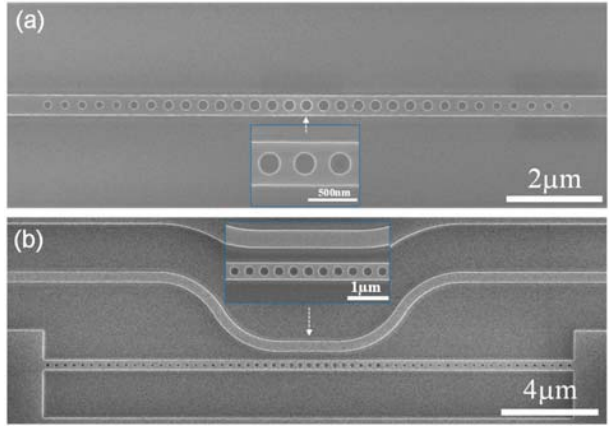

**Figure 4.** SEM images of fabricated 1D nanocavity. (a) Inline cavity and (b) waveguide-coupled cavity. The insets show the enlarged view of the center of cavities.

Firstly, to evaluate the quality of the cavities, we measured the transmission spectra of the inline 1D cavities and show the result in Fig. 5. The total quality factor $Q_{tot}$ of the cavities was extracted by fitting their resonance peaks with a Lorentzian lineshape. For the cavity with only 5 mirror holes, the fundamental mode has a $Q_{tot}$ of $1.52 \times 10^4$ with a transmittance of 95%. The addition of mirror holes can reduce the waveguide coupling loss and thus increase $Q_{tot}$. For instance, the cavity with 10 mirror holes has a $Q_{tot}$ of up to $1.68 \times 10^5$, simultaneously sustaining a considerable transmittance of 56%. Note that for the fundamental resonance peak the side-mode suppression ratio is as high as 40 dB. These characteristics are essential for narrow-band, low-insertion-loss, high-extinction-ratio filter designs. For the cavity with 15 mirror holes, the measured $Q_{tot}$ is $0.65 \times 10^6$, and by using $T(\lambda_{res}) = (Q_{tot}/Q_w)^2$ and Eq. (2) the intrinsic $Q_{int}$ of $0.83 \times 10^6$ is obtained. Since in our cavity $Q_{rad}$ is $>10^7$, the intrinsic $Q_{int}$ is mainly limited by scattering loss ($Q_{scat}$). By scaling the dimension of the cavity (e.g., waveguide width or hole radius), we are able to tune the fundamental resonance over the C-band, meanwhile maintaining similar high $Q$ factors, as shown in Fig. 6. We demonstrated the highest $Q_{tot}$ of $0.78 \times 10^6$ and the highest $Q_{int}$ of $0.84 \times 10^6$ among the measured cavities, a nearly 20 times improvement over previously demonstrated 1D nanocavities patterned by photolithography [20]. To the best of our knowledge, this is also the highest $Q$ factor demonstrated in 1D PhC nanocavities overall, even exceeding those measured in air-bridge cavities fabricated by EBL [6]. Such high *Q* factors enable us to realize more advanced PICs with higher degree of functionality as discussed below.

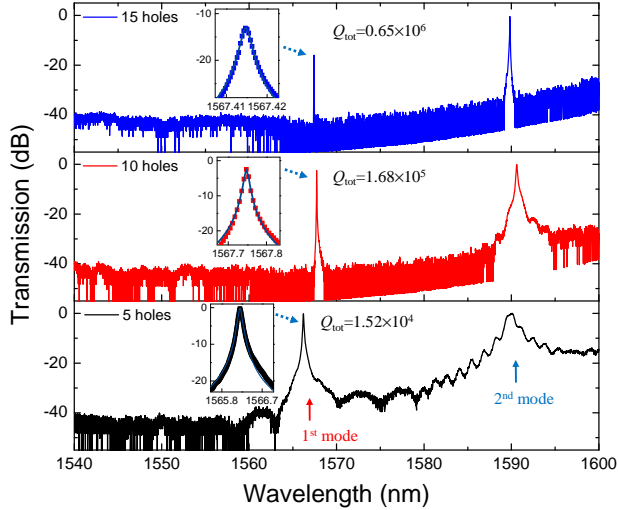

**Figure 5.** Normalized transmission spectra of inline 1D nanocavities with 5, 10, and 15 mirror holes respectively at each side of the cavity. The insets show the corresponding Lorentzian lineshape fittings for the fundamental (1st order) modes at about 1566 nm. Note that also the 2nd order modes are plotted, which are red-shifted with respect to the fundamental resonance by about 22.5 nm. The measured cavities have a nominal width of 456 nm.

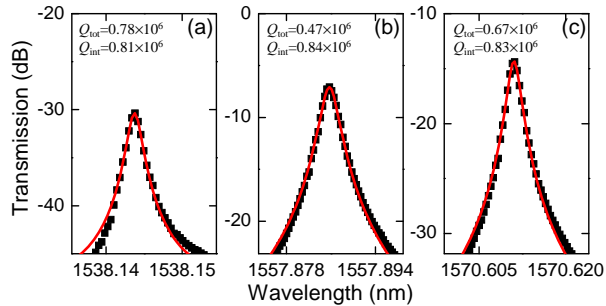

**Figure 6.** The measured and fitted *Q* factors of the fundamental modes for different cavities. The resonance wavelength is tuned over the C-band by varying the waveguide width and/or hole radius. The waveguide width and radius of the center hole are: (a) 438 nm and $1.05r_0$, (b) 456 nm and $1.05r_0$, and (c) 456 nm and $1.0r_0$, respectively, with $r_0$ = 110 nm as defined above.

Next, we studied the spectral response of the side-coupled nanocavity. Fig. 7(a) shows the transmission spectra at different coupling gaps. The measured transmission exhibits notch-filter-like characteristics, agreeing well with the theoretical analysis and numerical simulation. According to our coupled mode analysis (Eqs. (9) and (11)), given a certain intrinsic $Q_{\text{int}}$ for a cavity, $Q_{tot}$ and the transmission dip are determined by the coupling strength between waveguide and cavity. Thus by adjusting the coupling gap, we can tune $Q_{tot}$ and the transmission dip effectively, as clearly shown in Fig. 7(a). For a gap of 450 nm, $Q_{tot}$ is $8\times10^4$ and the transmission is suppressed by 16 dB. From Eq. (11), we then calculated $Q_{tot} \approx 0.16 Q_{int}$, which means that for this cavity the coupling loss to the bus waveguide is dominant. The coupling for high-order modes is significantly suppressed, thanks to the spatial separation of the cavity and waveguide modes. The 2$^{nd}$ mode shows only a 1.5 dB transmission dip at the gap of 250 nm while it disappears at larger gaps. This distinguishing feature enables single-wavelength operation over a very broad wavelength range, compared to conventional filters limited by free-spectral ranges (e.g., ring resonators). Using Eq. (11), we calculated $Q_{int} = 0.5\times10^6$ for the cavity with a gap of 450 nm, which is smaller than those of individual inline cavities. The reduction of $Q_{int}$ might be attributed to parasitic scattering loss due to the presence of the bus waveguide and possible fabrication imperfections due to proximity effects in the photolithography process.

By coupling a bus waveguide with a series of 1D nanocavities, we can also achieve multi-wavelength operation with arbitrary yet predefined resonance wavelengths for each cavity. To demonstrate this, we designed a structure with a bus waveguide successively coupled with five cavities whose central hole radii are scaled to shift the resonance wavelength. Fig. 7(b) shows the corresponding transmission spectrum with the five fundamental resonances clearly visible. The shift of the resonance wavelength vs. the change of the hole radius is shown in Fig. 7(c). On average the resonance is blue-shifted by 2.3 nm per 1.0 nm increment in radius. Note that for the cavities with larger radii the 2$^{nd}$ order modes are also coupled with considerable transmission dips. This can be understood by the fact that the higher-order modes are spatially more extended for devices with larger radius, leading to increased coupling with the waveguide mode. Eventually, the coupling for higher-order modes could be eliminated by improving the design. For instance, instead of changing the radius, we can shift the cavity resonance by adjusting other geometric parameters such as the width and lattice period or by scaling the structure as a whole.

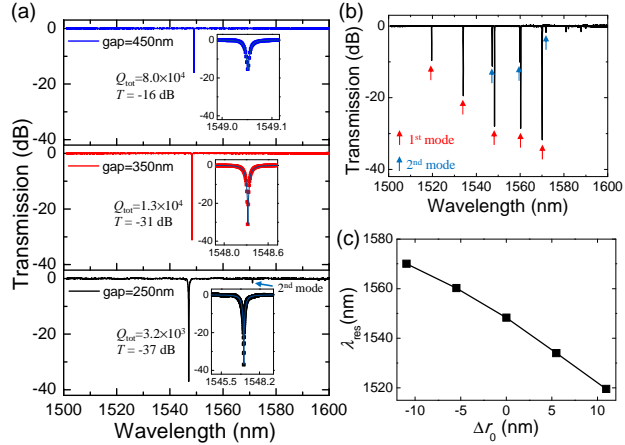

**Figure 7.** Transmission spectrum results of waveguide side-coupled nanocavities. (a) Normalized transmission spectra with coupling gaps of 250, 350, 450 nm. The insets show the corresponding fittings for the fundamental modes together with $Q_{tot}$ and the transmittance at the resonance (*T*). For the spectrum with the gap of 250 nm, the 2nd mode is very weakly coupled with a transmission dip of 1.5 dB, as indicated by the arrow. The measured cavities have a nominal width of 438 nm. (b) Normalized transmission spectrum of a single bus waveguide side-coupled with five nanocavities of which the radius of the central hole is linearly scaled from $0.9r_0$ to $1.1r_0$ ($\Delta r_0$ from -10.95 nm to +10.95 nm). All the cavities have the same width of 438 nm and the same mirror hole radius. The coupling gap is fixed at 350 nm for all cavities. The red and blue arrows indicate the fundamental (1st order) and the 2nd order modes respectively. (c) The measured fundamental resonances vs. the change of radius for all 5 cavities.

So far, we have focused on the transmission properties of the inline and waveguide-coupled nanocavities. In several cases, e.g., when used as a mirror in a laser cavity, also the reflection is important. Therefore, to verify our theoretical analysis and numerical simulation carried out above, we also studied the reflection spectra of the side-coupled nanocavities. A fiber circulator was used and the coupling fibers, as well as the whole device, were immersed in index-matching fluid. As a result, the reflection from the fiber facets could be significantly suppressed and the associated interference ripples in the reflection spectrum could be minimized. Note that the index-matching fluid red-shifts the resonances of the cavities by about 16 nm due to the cladding effect. We measured the same devices as described in Fig. 7(a) and show the reflection spectra in Fig. 8(a) together with the transmission spectra for comparison. Note that there are still ripples in the spectra due to residual reflection of the fiber facets and other reflection mechanisms such as reflection of gratings and substrate. Nevertheless, the resonance peaks can be clearly seen in the spectra with a local side-mode suppression ratio of more than 10 dB. Both the reflection and transmission spectra are in good agreement with our analytical and numerical results. When increasing the gap, the quality factor $Q_{tot}$ of the reflection peak increases from $2.1\times10^3$ to $4.6\times10^4$, while the reflectance decreases slowly from 95% to 80%, suggesting $Q_{tot}<<Q_{int}$ from Eq. (11), that is, the decay of the cavity mode into the waveguide dominates over its intrinsic loss at all gaps. We also measured the reflection spectrum of the waveguide coupled with multiple cavities and show the result in Fig. 8(b). As expected, the resonances of the different cavities are independently coupled out and clearly visible in the reflection spectrum of the bus waveguide. From Fig. 8, both the transmittance ($T$) and reflectance ($R$) at resonance can be obtained. As Eq. (11) results in $\sqrt{T}+\sqrt{R}=1$, we can use the experimental data to verify the analytical result quantitatively. Table 1 summarizes the calculated $\sqrt{T}+\sqrt{R}$ using the measured data for all 8 fundamental resonances in Fig. 8(a) and 8(b). The experimental results for all those peaks give $\sqrt{T}+\sqrt{R}\approx 1$, agreeing well with the analytical result. Therefore, our measurement, in turn, demonstrates that the theoretical model describes the waveguide-nanocavity coupled structure with very good accuracy, indicating it can be used in practical device designs.

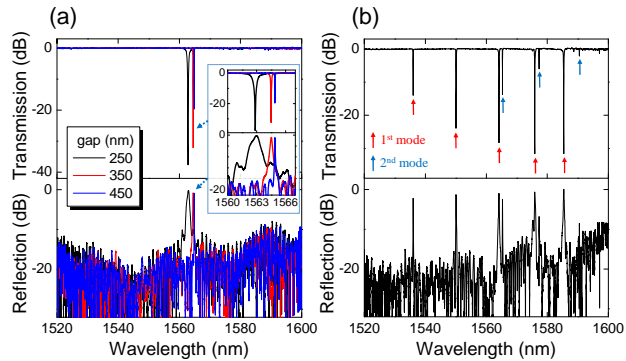

**Figure 8.** Reflection measurement results of waveguide side-coupled nanocavities. (a) Reflection and transmission spectra of the devices described in Fig. 7(a). The inset shows a close-up of the fundamental modes. (b) Reflection and transmission spectra of the multi-wavelength device measured in Fig. 7(b). The red and blue arrows indicate the 1st order and 2nd order modes, respectively.

**Table 1**. Calculated ($\sqrt{T} + \sqrt{R}$) for the 8 fundamental resonance peaks in Fig. 8

| Peak | $T$ | $R$ | $\sqrt{T} + \sqrt{R}$ |
|---|---|---|---|
| 1 | 0.01104 | 0.80724 | 1.00354 |
| 2 | 6.19E-04 | 0.82414 | 0.93271 |
| 3 | 1.75E-04 | 0.9506 | 0.98822 |
| 4 | 0.04064 | 0.60117 | 0.97696 |
| 5 | 0.00411 | 0.74302 | 0.92611 |
| 6 | 0.00152 | 0.79433 | 0.9302 |
| 7 | 6.98E-04 | 0.86696 | 0.95753 |
| 8 | 7.01E-04 | 0.99998 | 1.02644 |

## 4. Discussion and Summary

We have systematically studied the $Q$ factors of 1D PhC nanocavities fabricated using optical immersion lithography and experimentally demonstrated $Q$ factors of up to $0.84 \times 10^6$ in the C-band with a footprint of only 0.5µm × 22µm. Such compact ultrahigh-$Q$ cavities hold the promise of forming the basis of more complex photonic ICs relevant for various applications. In cases requiring low insertion loss and moderate $Q$ (e.g. $<10^4$), like modulators, we can design the cavities having a high transmittance of >97%. On the other hand, the same basic cavities also allow to realize high-extinction-ratio and narrow-band filter designs, for instance, a filter with a bandwidth of 0.8 GHz ($Q$ of ~$0.25 \times 10^6$) allows for a transmittance of 50%. While the inline devices are limited to single-wavelength operation, the bus-coupled cavities allow not only for both transmission and reflection-type operation but also for multi-wavelength operation. We have experimentally demonstrated both notch filters and wavelength selective reflectors with this architecture, and showed multiple wavelength operation over a broad wavelength range of more than 100 nm. We also clearly demonstrated, both theoretically and experimentally, that high intrinsic $Q$ factors are crucial for combining narrow band and low loss operation. For example, considering an intrinsic $Q$ factor of $0.5 \times 10^6$ in a bus-coupled nanocavity (as measured in our fabricated devices) and requiring a reflectance of ≥50%, the reflection bandwidth can be as narrow as ~1 GHz ($Q$ of ~$0.15 \times 10^6$), while the device length is only ~22 µm. For comparison, to achieve a similar bandwidth, for example, a distributed Bragg reflector with low grating strength needs centimeter-scale length [25]. Overall, the high $Q$-factor devices allow for flexible operation and a wide tunability in terms of bandwidth, transmittance and reflectance, simply by controlling the coupling gap. Alternatively, this tunability can be implemented by dynamic tuning of the cavity $Q$ using, for example, electro-optic effects. Lastly, due to their waveguide-coupled nature and compactness, the demonstrated building blocks can be readily incorporated into devices with a higher level of complexity, such as wavelength-division multiplexing modulators and tunable or narrow-linewidth lasers.

In conclusion, we have reported the realization of oxide embedded 1D PhC nanocavities fabricated with optical immersion lithography in a CMOS-pilot line with record $Q$ factors of up to $0.84 \times 10^6$. Starting from these basic cavities, we further investigated both theoretically and experimentally the properties of bus waveguide coupled structures and showed that this simple configuration allows to flexibly manipulate the spectral response both in transmission and reflection. We demonstrated this through the realization of narrow-band and high-efficiency notch filters and wavelength selective reflectors. In stark contrast with e.g., ring resonators, the demonstrated devices show single mode operation over a very large wavelength range, even extending beyond 100 nm. By coupling a bus waveguide with a series of cavities, we also achieved multi-wavelength operation in which both resonance wavelength and mode number can be controlled as desired. This efficient resonance management in those ultrahigh-$Q$ nanocavities opens up new opportunities for the design of compact, high-performance, resonance-based photonic components such as filters, reflectors, and so on. Those components will be potentially attractive in fundamental resonator-related researches and practical device applications like modulators, optical sensors, and complex lasers. Given their compatibility with advanced optical lithography and fabrication in a CMOS fab, our result paves the way for the implementation of such ultrahigh-$Q$ nanocavities in large-scale complex PIC systems.

**Funding**

This work was partly supported by the European Commission through the Horizon 2020 Framework Programme (Grant agreement No. 732894 - FET proactive HOT).

**Acknowledgements**

The authors acknowledge imec's industrial affiliation Optical I/O program.